\documentclass[aps,prd,amsmath,amssymb,twocolumn,preprintnumbers,nofootinbib]{revtex4}
\pdfoutput=1
\usepackage{graphicx}
\graphicspath{{./plots/}}
\usepackage{color}
\usepackage{hyperref}					
\usepackage{mathtools}					
\usepackage{slashed}					
\renewcommand{\epsilon}{\varepsilon}		

\newcommand{\beq}{\begin{eqnarray}}
\newcommand{\eeq}{\end{eqnarray}}

\newcommand{\bmp}{\noindent\begin{minipage}{16cm}}
\newcommand{\emp}{\end{minipage}\vskip 7mm} 

\def\simge{\mathrel{%
   \rlap{\raise 0.511ex \hbox{$>$}}{\lower 0.511ex \hbox{$\sim$}}}}
\def\simle{\mathrel{
   \rlap{\raise 0.511ex \hbox{$<$}}{\lower 0.511ex \hbox{$\sim$}}}}

\usepackage{dcolumn}
\usepackage{bm}
\usepackage{bbm}
\usepackage{subfigure}
\usepackage{pxfonts}

\definecolor{rossoCP3}{cmyk}{0,.88,.77,.40}

\def\lsim{\mathrel{\rlap{\lower4pt\hbox{\hskip1pt$\sim$}}
    \raise1pt\hbox{$<$}}}                
\def\gsim{\mathrel{\rlap{\lower4pt\hbox{\hskip1pt$\sim$}}
    \raise1pt\hbox{$>$}}}                

\newcommand{\be}{\begin{eqnarray}}
\newcommand{\ee}{\end{eqnarray}}

\begin{document}
 \includegraphics[width=2.8cm]{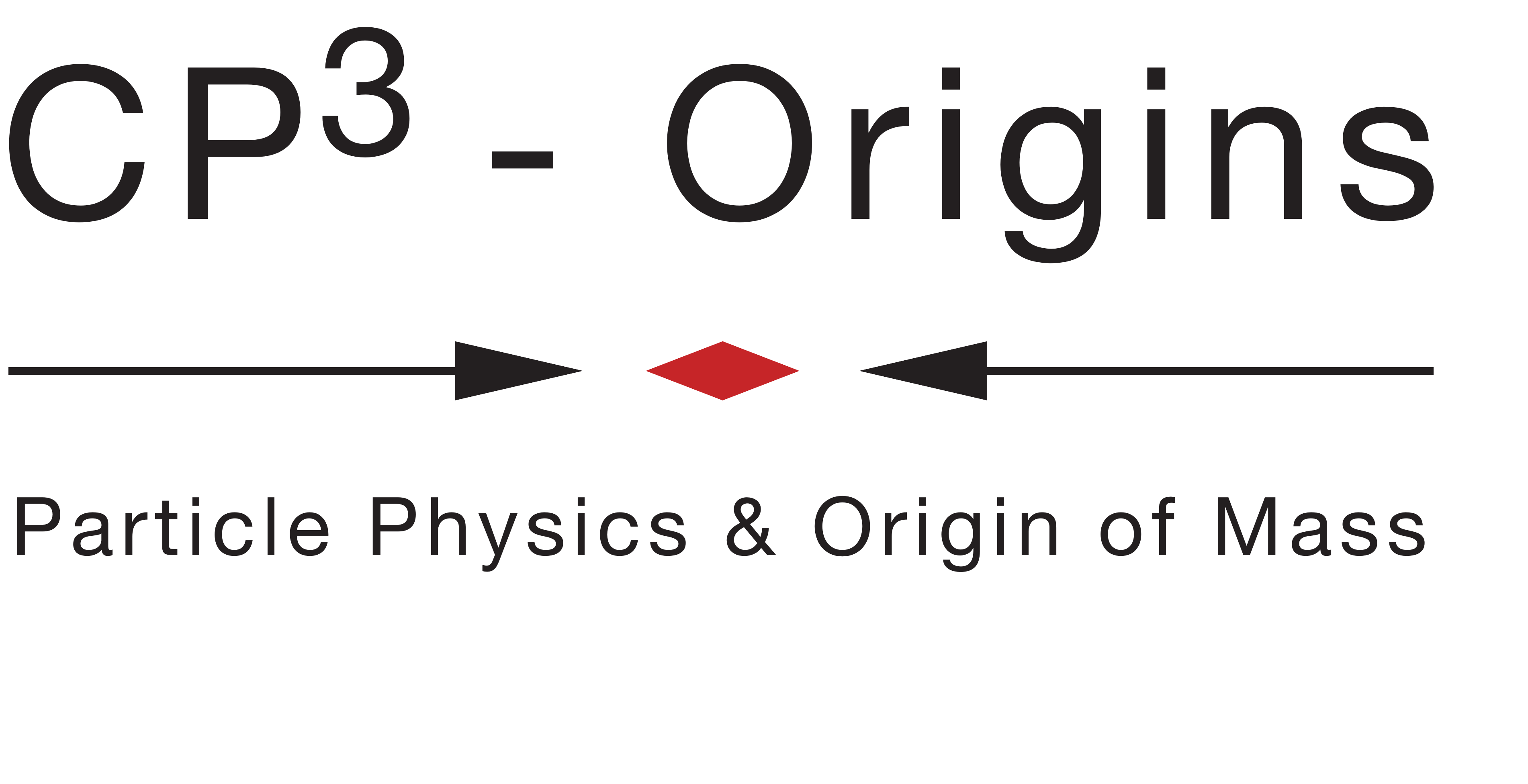}   
 \title{\Large  \color{rossoCP3} Dark Medium Modified Dispersion Relations}
\author{Isabella {\sc Masina}$^{\color{rossoCP3}{\varheartsuit}}$$^{\color{rossoCP3}{\clubsuit}}$ }
\email{masina@fe.infn.it}
\author{Francesco {\sc Sannino}$^{\color{rossoCP3}{\varheartsuit}}$ }
\email{sannino@cp3-origins.net} 
\affiliation{ 
$^{\color{rossoCP3}{\varheartsuit}}$
{ \rm CP}$^{ \bf 3}${\rm-Origins} \& the Danish Institute for Advanced Study { \rm DIAS},\\ 
{University of Southern Denmark, Campusvej 55, DK-5230 Odense M, Denmark} \\
$^{\color{rossoCP3}{\clubsuit}}$ 
{Dip.~di Fisica dell'Universit\`a di Ferrara and INFN Sez.~di Ferrara, Via Saragat 1, I-44100 Ferrara, Italy}}

\begin{abstract}
We suggest  that the dark side of the universe builds a dark medium modifying the Standard Model particle dispersion relations.  We introduce the generic form for the modified dispersion relations and provide explicit models of the dark medium. 
{We use the derived neutrino dispersion relations to show that the size of the corrections are smaller than the SN1987a bounds. 
We argue that before invoking modifications of Einstein's theory of relativity one should investigate the dark medium effects.  } 
\\ 
 [.1cm]
 {\footnotesize  \it Preprint: CP$^3$-Origins-2011-30  \& DIAS-2011-22} 
 \end{abstract}

\maketitle

 \section{Introduction}
\label{introduction}

The dispersion relations for the Standard Model (SM) particles are currently under experimental investigation. 
In particular due to the fact that neutrinos couple very weakly to matter they constitute an excellent case study of potential Lorentz violation. 
It is a well known fact, however, that matter effects can also deform the dispersion relations without invoking modification of Einstein's special relativity theory. 
This is, indeed, already the case for neutrinos propagating through Earth \cite{Wolfenstein:1977ue,Wolfenstein:1979ni,Mikheev:1986wj,Mikheev:1986gs}. 



On the other hand 
we are surrounded by a dark medium and therefore we investigate its potential effects on SM particles dispersion relations. 
We start by introducing the properties of the dark medium and introduce the generic form of the, in dark medium, modified dispersion relations for SM fermions. We provide explicit models of the dark medium featuring {\it dark gauge sectors} and use them to determine the changes in the SM fermions dispersion relations. 
We finally compare the results with experiments and show that the size of the corrections is much smaller than the SN1987a  experimental bounds \cite{Longo:1987ub}. 

\section{Dark medium and models}

Observations have demonstrated that our universe contains a baryonic medium providing a net matter density, it is therefore reasonable to assume that the dark side, holding the majority share of the universe, also builds a dark medium. Whatever this medium is constituted of, if the dark side couples to the SM fields can lead to in {\it dark medium} modifications of the SM particle dispersion relations. Here we will concentrate on the generic form of the dispersion relations for a SM fermion such as the neutrino. 

Let't start by defining with $u^{\alpha}$ the four-velocity of the center-mass frame of the dark plasma with 
$u^{\alpha} u_{\alpha} =1$. We assume the dark sector to contribute to the real part of the self-energy of a generic massless fermion according to \cite{Weldon:1982bn}:
\begin{eqnarray}
\Re \Sigma(K) = - a\,  \slashed{K}  - b \slashed{u} \ ,
\end{eqnarray}
with $a$ and $b$ functions of $K^2$ and $K\cdot u$. It is convenient to introduce the Lorentz scalars $\omega \equiv K\cdot u$ and $k \equiv  \sqrt{(K\cdot u )^2 - K^2}$. In the plasma frame, $u^{\alpha} = (1,0,0,0)$ and $\omega$ corresponds to the energy of the particle with respect to that frame and $k$ its momentum. The fermionic propagator becomes: 
\begin{eqnarray}
S(\omega,k) = \frac{1}{\slashed{K} - \Re \Sigma(K)} = \frac{(1+a)\slashed{K}+b \slashed{u}}{\left[(1+a)\omega +b\right]^2-\left[(1+a)k\right]^2} \ .
\end{eqnarray} 
Notice that $b(\omega,k)$ has dimension of energy, while $a(\omega,k)$ is dimensionless. The fermionic dispersion relations are obtained studying the zeros of the propagator \cite{Weldon:1982bn}.
It is always possible to define, once the solutions are determined for a specific theory, an index $n(k)$ as follows: 
\beq
\omega = k \,n(k)\,\ .
\eeq
Notice that $n(k)$ is the inverse of the refraction index, namely it is equal to the phase velocity. 
The group velocity is instead
\beq
v_g=\frac{\partial \omega}{\partial k}=n+k \frac{\partial n}{\partial k}\,\,.
\eeq
We consider the quasi-particle solution which is smoothly connected to the in vacuum solution when the $b$ term vanishes. In general there are four independent solutions. A detailed discussion can be found in \cite{Blaizot:1993bb}. 

If the medium violates parity the dispersion relations for the left and the right handed quasi-particles decouple \cite{Weldon:1982bn}:
\begin{eqnarray}
\left[(1+a_L)\omega_L +b_L\right]^2-\left[(1+a_L)k\right]^2 & = & 0 \ , \\ 
\left[(1+a_R)\omega_R +b_R\right]^2-\left[(1+a_R)k\right]^2 & = & 0 \ .
\end{eqnarray}
In the model section we will introduce simple models of dark medium. To compare with current experiments we assume that the particle investigated have a specific quasi-particle dispersion relation differentiating left from right. 
{In this case one has two independent  indices, $n_L$ and $n_R$. The left handed particle and its right handed anti-particle can travel at different speeds with respect
to the right handed particle and its left handed anti-particle. However, strictly speaking the SM contains only the left handed neutrino and the associated right handed anti-neutrino. Their index will be simply indicated by $n(k)$ in the following. 
 
 We consider, as a model,  the dark medium to be composed by a non-abelian gauge theory under which the ordinary neutrinos transform nontrivially. We will also assume that the dark particles, and the neutrinos, have non-zero chemical potentials.  The finite temperature and chemical potential corrections, leading to nonzero $a$ and $b$ functions,  for the asymptotic large $k$ behavior \cite{Morales:1999ia,Weldon:1982bn,Blaizot:1993bb}, $M\ll k$, are: 
\begin{eqnarray}
\omega(k) = k + \frac{M^2}{k} + {\cal O}(k^{-3}) \ ,  \quad n(k) = 1 + \frac{M^2}{k^2} \ ,
\label{sl}
\end{eqnarray}
with 
\begin{eqnarray}
M^2 = \frac{g^2 C(R)}{8} \left(T^2 + \frac{\mu^2}{\pi^2} \right) \ .
\end{eqnarray}
$g$ is the dark gauge coupling and $C(R)$ is the quadratic Casimir defined by $(T^A T^A)_{mn} = C(R) \delta_{mn}$, with $R$ the dark representation under which the neutrino transform. $A=1,...,N^2_D -1$ since we assume, for definitiveness,   $SU(N_D)$ to be the dark gauge group. 
 We consider the zero temperature case and are working in the dark plasma rest frame. 
%
%
The group velocity becomes
\beq
1-v_g =\frac{M^2}{k^2}=n(k)-1\,\,,
\eeq
so the propagation is subluminal. One realizes that the particle speed is slowed when the temperature is high and/or the chemical potential is significantly large.

From \eqref{sl} we observe a natural subluminal medium effect not due to an intrinsic Lorentz violation. One can use an Abelian dark medium yielding similar physical results \cite{Weldon:1982bn,Blaizot:1993bb}. Of course, there are many potentially interesting differences between an abelian and a non-abelian plasma such as the onset of self-induced superconductivity or superfluidity for the non-abelian case, which will be explored elsewhere.  


\section{Phenomenology of dark medium}

We now focus on the dark medium modified neutrino dispersion relations and, using \eqref{sl} as reference, we provide an estimate of the model deviations from the speed of light to be confronted with the experiments: 
\begin{equation}
n(k) - 1 \le  \left(\frac{m_{\nu}}{k}\right)^2  \ .
 \label{simple}
\end{equation}
In this expression we replaced $M$ by the neutrino mass $m_\nu$ which acts as a phenomenological upper limit for $M$ itself. 
The reason behind this choice is that, in the small $k$ limit and if the neutrinos studied originate on Earth, $M$ becomes the effective neutrino mass, and therefore cannot exceed the upper limit set by known experiments such as the tritium beta decay \cite{Nakamura:2010zzi}. 

Assuming $m_\nu \lesssim $ eV and taking  $k\sim 10$ MeV, we estimate $n(k) - 1  \lesssim {\cal O}(10^{-14})$  for electron anti-neutrinos coming from the explosion of the SN1987a \cite{Bionta:1987qt,Hirata:1987hu}. In the latter estimate, we used as upper limit for $M$ the mass of the neutrino which, in this case, can be replaced by the number density of the dark medium. 
To be concrete we use the number density for dark matter and write: 
\begin{equation}
 n(k) - 1 \le  \left(\frac{g^2}{8} \frac{N_D^2 - 1}{2N_D}\right) \left(\frac{1}{k}\right)^2  \left(  \frac{3}{\pi}  \frac{\rho_{DM}}{m_{DM}}    \right)^{2/3}  \ .
\end{equation}
The dark medium is composed by an $SU(N_D)$ gauge theory with dark gauge coupling $g$ and with fermions transforming according to the fundamental representation. This analysis applies immediately  to a dark abelian medium.  
We assume that the product of the gauge coupling times the quadratic Casimir over $8$ is around unity;
consistency of perturbation theory could require a further suppression of this estimate. Since the dark matter number density depends on the distance from the center
of the galaxy, the modification of the neutrino dispersion relations also depends on such distance. For definiteness, we focus on our galaxy.
The dark matter halo profile of the Milky Way can be modeled as follows: 
\beq
\rho_{DM}(r)= \rho_0 \frac{1}{(r/r_c)^{\gamma} [1+(r/r_c)^\alpha]^{\frac{\beta-\gamma}{\alpha}}} \ .
\eeq
 In the case of the Moore \cite{Moore:1999gc} profile we have $\alpha=1.5$, $\beta=3$, $\gamma=1.5$, $r_c=28$ kpc, $\rho_0=0.0585 {\rm GeV}/{\rm cm}^3$. 
 For the Navarro Frenk and White (NFW) \cite{Navarro:1995iw}  profile we have $\alpha=1$, $\beta=3$, $\gamma=1$, $r_c=20$ kpc and $\rho_0=0.26 {\rm GeV}/{\rm cm}^3$,
 while for the Isothermal case $\alpha=2$, $\beta=2$, $\gamma=0$, $r_c=3.5$ kpc, $\rho_0=2.08 {\rm GeV}/{\rm cm}^3$. 
 The profiles are normalized in order to have $\rho_{DM}(r_{sun})=0.30 {\rm GeV}/{\rm cm}^3$, at $r_{sun}= 8.5$ kpc. 
 The corresponding $n(k)-1$ profiles  are plotted in Fig.~\ref{nprof} as a function of the distance from the centre of our galaxy, for $k=10$ MeV and 
  $m_{DM} \sim m_{\nu} \simeq 1~{\rm eV}$  (upper curves) and $m_{DM}\simeq 10~{\rm GeV}$ (lower curves). 
 Within each set of curves the up most corresponds to the Moore profile, the middle to the NFW profile and the bottom to the Isothermal one. 
From Fig.~\ref{nprof}, obtained for $k=10~$MeV, emerges that, unless the $m_{DM}$ is extremely small, the estimate for SN1987a from \eqref{simple} is substantially larger than the one obtained here, and therefore this estimate can be seen as an upper bound. 

\begin{figure}[t!]
\includegraphics[width=8cm]{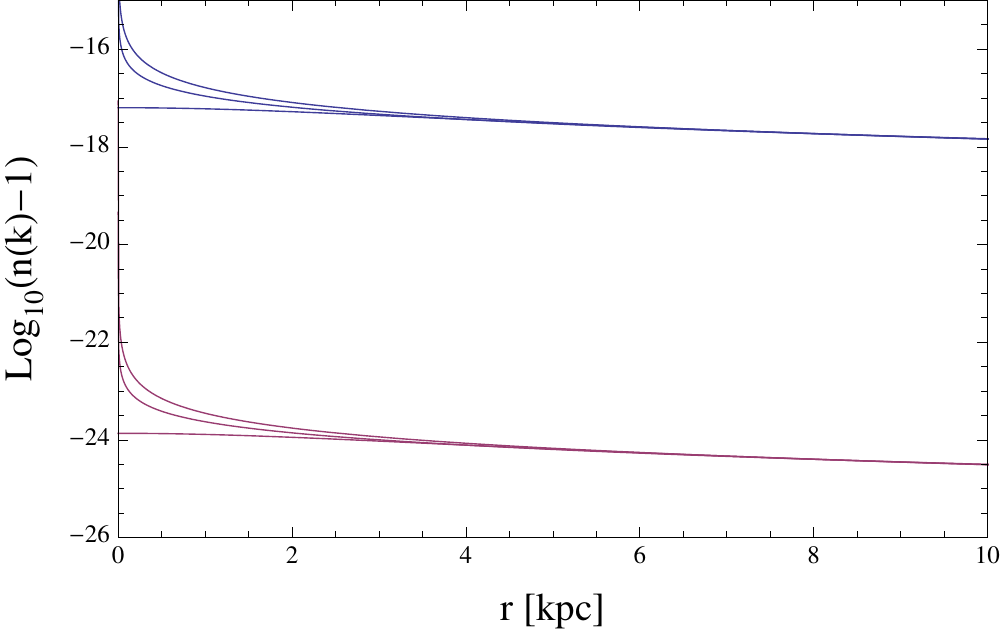}
\caption{ 
$n(k) - 1$ profile in the Milky Way galaxy for $k=10$~MeV. The upper curve is obtained for $m_{DM} \sim m_{\nu} \simeq 1~{\rm eV}$ while the lower curve is obtained for $m_{DM}\simeq 10~{\rm GeV}$. On the $x$-axis we report the distance from the center of the Milky Way.}
\label{nprof}
\vskip .2cm
\end{figure}

Our predictions are in full agreement with the SN1987a constraints on the neutrino propagation.
The SN1987a emitted all neutrino and anti-neutrino flavors, but only the electron anti-neutrinos could be detected by the experiments running in 1987.
The observation that the photons and the electron anti-neutrinos emitted by the SN1987a arrived within a few hours actually implies 
$(c-v_{\bar \nu_e})/c=n_{\bar \nu_e}(k) -1\lesssim 10^{-9}$  for $k ={\cal O}(10)$ MeV \cite{Longo:1987ub}. 
The fact that all the detected electron anti-neutrinos arrived within $10$ seconds,
is naturally accounted for if $n_{\bar \nu_e}(k) -1\lesssim 2\times 10^{-12}$  at $k \approx 10$ MeV. The latter bound corresponds to a sensitivity to $M$ at the
level of about  $15$ eV. Since the value of the effective mass $M$ is naturally smaller than the eV-scale, we can conclude that dark medium effects induced
a time spreading in the SN1987a neutrinos which was smaller than $ 0.1$ seconds, hence safely negligible.

\section{Conclusions} 
We investigated the dark medium effects on the SM particle dispersion relations. 
We introduced the generic form for the modified dispersion relations and provided explicit models of the dark medium to compute them.  
In these models the fermions acquired speeds smaller than light in the vacuum. 
In general we expect, and our models support this expectation, the dark medium effects to disappear at very large momenta 
given that almost any ordinary medium becomes transparent in this regime. 
We have then shown that with natural values for the dark medium densities the size of the corrections to the neutrino dispersion relations 
are much smaller than the bounds set by SN1987a. It is worth mentioning that in \cite{Sannino:2003mt,Sannino:2003ai} the modified 
dispersion relations for the SM fermions were predicted assuming a Bose-Einstein (BE) relativistic condensation in the bosonic sector of the SM, 
or in a simple extension of it. The corrections are similar to the ones presented here.

{ It is relevant to mention that we expect that it is possible to disentangle modifications to the dispersion relations due to, in medium effects, from the ones coming from, for example, an explicit breaking of Lorentz violation coming from a more complete gravitational theory. The reason being that, in the second case, the dispersion relations should receive corrections inversely proportional to an high energy scale such as the Planck scale or any other ultraviolet scale. This implies that the gravitational corrections to the dispersion relations are larger at higher energies while the ones coming from, in medium corrections, disappear at very high energies. }

The size of the dark medium corrections to Lorentz preserving SM particle dispersion relations, clearly shows that these effects should 
be accounted for, before invoking drastic modifications of Einstein's theory of relativity. 
 
\section*{Acknowledgements} We thank A.Strumia for useful discussions.


\onecolumngrid
\vfill \hfill
  \includegraphics[width=2.cm]{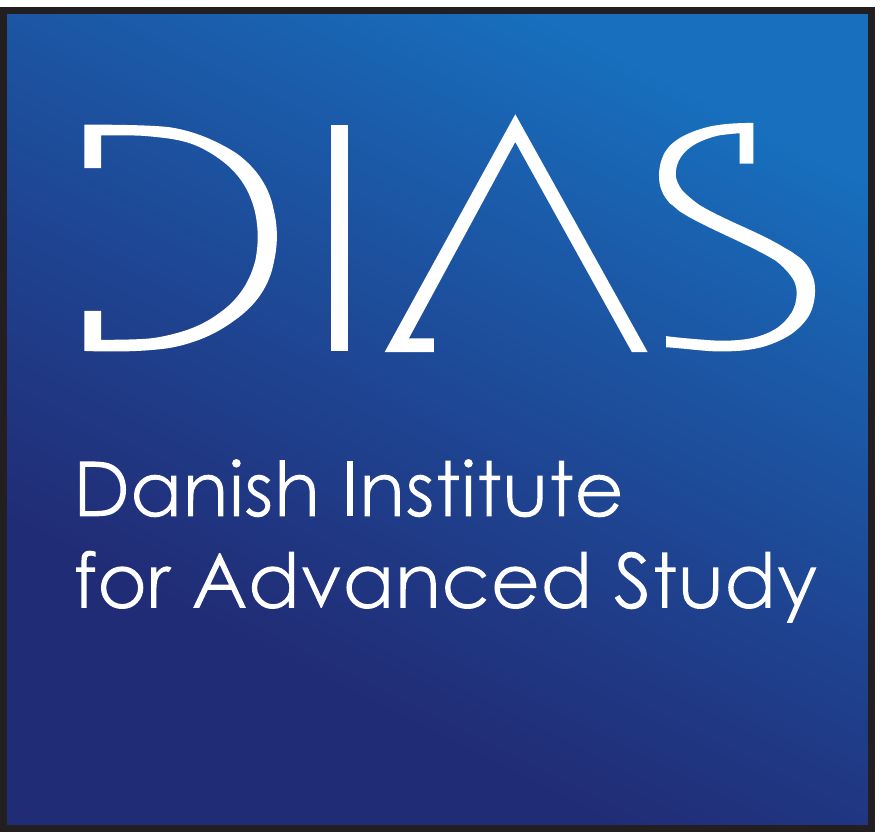}}
 \end{document}